\documentclass[a4paper]{article}
\usepackage{amssymb}
\usepackage{graphicx}

\newcommand{\R}{\mathbb{R}}
\newcommand{\N}{\mathbb{N}}

\title{\large{{\bf Does Newtonian dynamics need Euclidean space?}}}

\author{}

\date{}

\begin{document}
	
	\maketitle

\centerline{Alain Albouy}
\centerline{CNRS, Observatoire de Paris}
\centerline{61, avenue de l'Observatoire}
\centerline{F-75014 Paris}

\bigskip
\bigskip
\bigskip

The motion of a ``planet'' of negligible mass around a fixed ``Sun'' of mass $m>0$, at the origin ${\rm O}$ of the space ${\rm O} xyz=\R^3$, is described by the vector differential equation
$$\pmatrix{\ddot x\cr \ddot y\cr \ddot z}=-\frac{m}{r^3}\pmatrix{x\cr y\cr z},\quad\hbox{with }r=\sqrt{x^2+y^2+z^2}.\eqno(N)$$
We use standard notation: the real numbers $x$, $y$, $z$ are the coordinates of the planet, and form a column vector. The double dot denotes the second derivative with respect to the time $t$. The left-hand side is the acceleration vector. We have removed the gravitational constant from the right-hand side by choosing the physical units of time, length and mass in such a way that this constant is 1.

Does Newtonian dynamics need a Euclidean space? System $(N)$ is defined on a 3-dimensional vector space. A trajectory may be seen as a curve in the position space together with a velocity vector at each position. The acceleration vector is the time derivative of the velocity vector. This derivative may be computed because the velocity vectors at different positions belong to the same vector space, namely, the vector space ${\rm O} xyz$. This is a property of a vector space, classically stated as: at any point the tangent space of a vector space is the vector space itself (in contrast, the tangent space to a sphere at a point is a plane which depends on the point). So, yes, Newtonian dynamics needs a vector space.

Another ingredient in $(N)$ is the Euclidean distance $r$ of the planet to the origin.  In conclusion, yes, $(N)$ is formulated in a Euclidean vector space. But, does $(N)$ possess generalizations in other types of spaces? We would like to have not only the same kind of equation but also the same kind of solutions.

Call a solution {\it radial} whenever the planet remains on a ray starting at ${\rm O}$. The bounded nonradial solutions of system $(N)$ are described by Kepler's laws.

$({\rm K}1)$ The planet moves on an ellipse with the Sun at a focus.

$({\rm K}2)$ On a given orbit, equal areas are swept out in equal times.

$({\rm K}3)$ The square of the period is proportional to the cube of the semi major axis.

\smallskip

Note that $({\rm K}1)$ states in particular that the orbit is planar. {\it We will only study the 2-dimensional version of $(N)$, or, in other words, we will state once and for all that $z=0$}.

 The law $({\rm K}2)$ defines a proportionality factor $C$, called the areal velocity or the angular momentum. A complete characterization of the motions should give the value of $C$ on each orbit. The law $({\rm K}3)$ indirectly does this, as we will check later. However, it introduces a new proportionality factor but does not give its value. Kepler claimed that this factor is the same for all the planets orbiting around the Sun. If the period $T$ is counted in years, and the semi major axis is counted in astronomical units, this factor is 1, since this is its value for the planet Earth. Consider now Newton's equation $(N)$, which was not known to Kepler. The law $({\rm K}3)$ should be stated with an explicit proportionality factor for the solutions of $(N)$. Let us do it. Let  $n=2\pi/T$ be the angular frequency, and $a$ the semi major axis. The precise Newtonian version of $({\rm K}3)$ is $n^2a^3=m$, where $m$ is the mass of the Sun. The bounded nonradial solutions of $(N)$ are now perfectly described.

The mathematical proof of the compatibility of equation $(N)$ with Kepler's law was published by Newton in 1687, in his {\it Principia} \cite{New}. This equivalence of a geometrical description and a description in terms of a differential equation remains a jewel of mathematical physics. It played an extraordinarily important role in the development of science. 

Many proofs followed Newton's proof, sometimes elegant and short. However, we would expect that, today, a student with good training in geometry and analysis could prove this equivalence on a first attempt. All the proofs I know require tricks which must be learned, as guessing them is far from easy. Well-known tricks are replacing the variable $r$ by the variable $1/r$ and the time $t$ by the polar angle $\theta$.

I will, as Newton did in the beginning of his {\it Principia}, start from Kepler's law and deduce the Newtonian acceleration $(N)$. This deduction will not require any tricks. By going backward along the same path, we will obtain Kepler from Newton. This reversed path is a known and extremely useful method. The novelty of my presentation is thus mainly pedagogical. But it will lead us directly to some less well-known generalizations and remarks.

\bigskip

\centerline{\bf 1. Kepler implies Newton in a simple way}

\bigskip

We all like to describe ellipses using a pair of foci. But Kepler's first law only mentions one focus. While devising the proof that Kepler's laws imply $(N)$, we would prefer a presentation of the conic sections reported by Pappus, where, instead of two foci, a focus and the associated directrix are given. Observe this advantage of giving a directrix rather than a second focus: when the ellipse becomes a parabola, the second focus goes to infinity, while the directrix does not. We will call the equation $$r=\alpha x+\beta y+p,\quad\hbox{with }(\alpha,\beta,p)\in \R^3,\hbox{ } p>0\eqno(P)$$ the {\it focus-directrix equation} of a conic section with focus at the origin. The directrix is the line with equation $0=\alpha x+\beta y+p$. The equation $(P)$ expresses that the distance of the point $(x,y)$ to the directrix is proportional to its distance $r$ to the focus.

Observe this advantage of giving $(\alpha,\beta,p)$ rather than the directrix: in the circular case $\alpha=\beta=0$, the directrix is undefined. Note also that the useless branch of the hyperbola is excluded. We have the perfect equation for the planar orbits and we can now begin the proof. The direction of the velocity vector is obtained by differentiating $(P)$ with respect to time:
$$\Bigl(\frac{\partial r}{\partial x}-\alpha\Bigr)\dot x+\Bigl(\frac{\partial r}{\partial y}-\beta\Bigr)\dot y=0.$$
So, for some real quantity $\lambda$, the velocity vector has coordinates $$\dot x=-\lambda\Bigl(\frac{\partial r}{\partial y}-\beta\Bigr), \qquad \dot y=\lambda\Bigl(\frac{\partial r}{\partial x}-\alpha\Bigr).$$
Now Kepler's second law should give some information about $\lambda$. Let $A$ be the area swept out from ${\rm O}$ during the time interval $[t_0,t]$. Let $C$ be the areal velocity: The formula corresponding to the law $({\rm K}2)$ is $2A=C(t-t_0)$. Then
$$C=2\dot A=x\dot y-y\dot x=\lambda x\Bigl(\frac{\partial r}{\partial x}-\alpha\Bigr)+\lambda y\Bigl(
\frac{\partial r}{\partial y}-\beta\Bigr)=\lambda(r-\alpha x-\beta y)=\lambda p.$$ Here we use $r=x\partial r/\partial x+y\partial r/\partial y$ which expresses the homogeneity of the function $r$. The proportionality factor $\lambda$ appears to be constant along an orbit. Since we assumed $p\neq 0$
$$\dot x=-\frac{C}{p}\Bigl(\frac{\partial r}{\partial y}-\beta\Bigr), \qquad \dot y=\frac{C}{p}\Bigl(\frac{\partial r}{\partial x}-\alpha\Bigr).\eqno(V)$$
Differentiating again with respect to time we get the coordinates of the acceleration vector
$$\ddot x=-\frac{C}{p}\Bigl(\frac{\partial^2 r}{\partial y\partial x}\dot x
+\frac{\partial^2 r}{\partial y^2}\dot y\Bigr), \qquad \ddot y=\frac{C}{p}\Bigl(\frac{\partial^2 r}{\partial x^2}\dot x
+\frac{\partial^2 r}{\partial x\partial y}\dot y\Bigr).\eqno(S)$$
But $$\frac{\partial r}{\partial x}=\frac{x}{r},\quad\frac{\partial r}{\partial y}=\frac{y}{r},\quad\frac{\partial^2 r}{\partial x^2}=\frac{y^2}{r^3},\quad \frac{\partial^2 r}{\partial x\partial y}=-\frac{xy}{r^3},\quad\frac{\partial^2 r}{\partial y^2}=\frac{x^2}{r^3}.\eqno(D)$$
So $$\ddot x=-\frac{C}{p r^3}\bigl(-xy\dot x
+x^2\dot y\bigr)=-\frac{C^2x}{p r^3}, \qquad \ddot y=\frac{C}{p r^3}\bigl(y^2\dot x
-xy\dot y\bigr)=-\frac{C^2y}{p r^3}.\eqno(T)$$
The acceleration appears to be central. But still the expression $C^2/p$ should be determined. Since $\pi ab$ is the area of the ellipse, where $b$ is the semi minor axis, the second law $C(t-t_0)=2A$ applied to a complete revolution gives $CT=2\pi a b$. This is also $C=nab$, where $n=2\pi/T$.  The coefficient $p$ of the focus-directrix equation is the {\it semi-parameter}. It satisfies the classical relation $b^2=ap$. Then $C^2/p=n^2a^2b^2/p=n^2a^3=m$, since Kepler's third law is $n^2a^3=m$.
Replacing this in the previous equation gives the Newtonian attraction $(N)$. We straightforwardly deduced the Newtonian attraction from Kepler's three laws.

\bigskip

\centerline{\bf 2. An astonishing generalization}

\bigskip

A careful analysis of the above deduction suggests starting again from the beginning. Elaborating on a result by Jacobi \cite{Jac} and on remarks by Darboux \cite{Dar}, we state three {\it Kepler-Jacobi's laws}.

$({\rm K}1')$ The planet moves on a $\rho$-focus-directrix curve with the Sun at the focus.

$({\rm K}2)$ On a given orbit, equal areas are swept out in equal times.

$({\rm K}3')$ The quotient of the square of the areal velocity $C^2$ by the semi-parameter $p$ is the mass $m$ of the Sun.

\smallskip\noindent A $\rho$-focus-directrix curve is defined by a function $\rho:{\rm O} xy\setminus \{{\rm O}\}\to\R$ and by three real parameters. We assume that $\rho$ is differentiable enough and {\it positively homogeneous of degree 1}, that is, satisfies $\rho(\lambda q)=\lambda\rho(q)$ for any $q\in {\rm O} xy\setminus \{{\rm O}\}$ and any $\lambda>0$. A $\rho$-focus-directrix curve is a curve with equation
$$\rho=\alpha x+\beta y+p,\quad\hbox{with }(\alpha,\beta,p)\in \R^3,\hbox{ } p>0.\eqno(P')$$
Focus and directrix are defined as in the case $\rho=\sqrt{x^2+y^2}$, and $p$ is still called the semi-parameter. We should now compute the Newton-Jacobi acceleration, and the computation goes exactly as in \S 1, up to equation $(S)$. To get the analog of $(D)$, we look at the Hessian matrix
$$\partial^2 \rho=\pmatrix{\partial^2 \rho\over \partial x^2&\partial^2 \rho\over \partial x\partial y\cr \partial^2 \rho\over \partial y\partial x&\partial^2 \rho\over \partial y^2}.$$
Both functions $\partial \rho/\partial x$ and $\partial \rho/\partial y$ are positively homogeneous of degree 0. The corresponding Euler equations show that the vector $$\pmatrix{x\cr y}$$ belongs to the kernel of $\partial^2 \rho$. This symmetric matrix is consequently determined up to a factor. There is a function $\rho^{\{2\}}:{\rm O} xy\setminus \{{\rm O}\}\to \R$ such that
$$\partial^2 \rho=\rho^{\{2\}}\pmatrix{y^2&-xy\cr -xy&x^2}.\eqno(H)$$
For example, the function $r=\sqrt{x^2+y^2}$ gives, according to formulas $(D)$,
$$\partial^2r=\frac{1}{r^3}\pmatrix{y^2&-xy\cr -xy&x^2},$$
which means $r^{\{2\}}=1/r^3$.
For a general function $\rho$ of same homogeneity as $r$, the proportionality factor $\rho^{\{2\}}$ is still positively homogeneous of degree $-3$. We get instead of $(T)$:
$$\ddot x=-\frac{C^2}{p}x\rho^{\{2\}}, \qquad \ddot y=-\frac{C^2}{p}y\rho^{\{2\}}.$$
Now the Kepler-Jacobi law $({\rm K}3')$ gives the Newton-Jacobi equation
$$\pmatrix{\ddot x\cr \ddot y}=-m\rho^{\{2\}}\pmatrix{x\cr y}.\eqno(N')$$
{\it We obtain a central force which does not depend on the velocity.} The expression for $\rho^{\{2\}}$ is elementarily computed as soon as $\rho$ is given. 
Let us take for example $\rho=\root 4 \of {x^4+y^4}$. Then $\rho^{\{2\}}=3x^2y^2(x^4+y^4)^{-7/4}$.
In this example we observe that the force vanishes along the coordinate axes. The orbit has an inflection point of order two when it crosses an axis (see Figure 1).

 \bigskip
\centerline{\includegraphics[width=60mm]{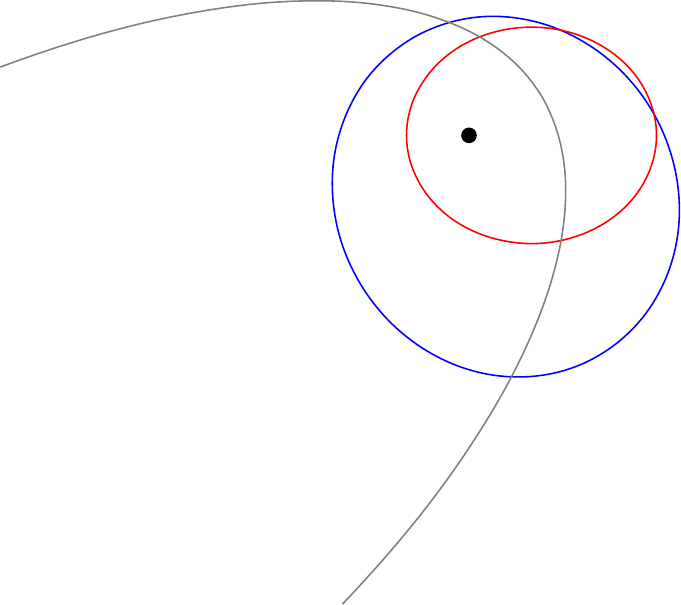}\qquad\includegraphics[width=60mm]{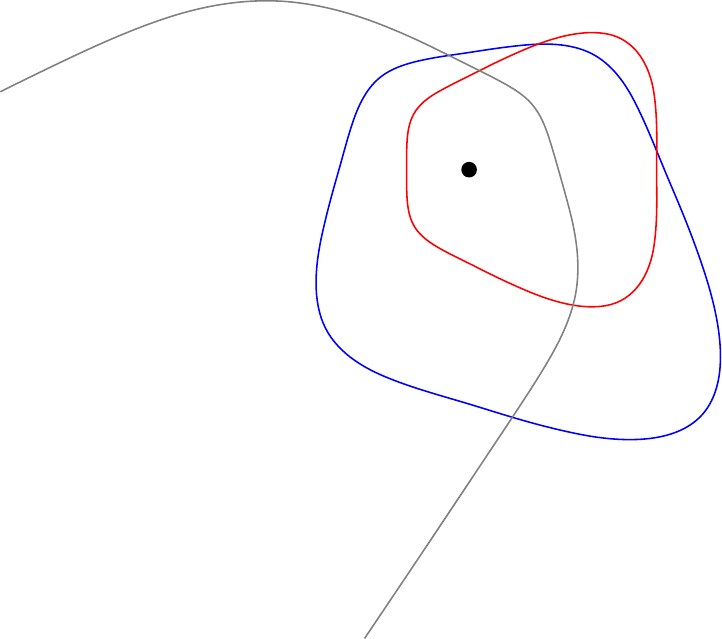}}

\nobreak
\centerline{Figure 1. Some Keplerian orbits and some orbits for $\rho^4=x^4+y^4$}
\bigskip

In the Keplerian case $({\rm K}3')$ describes the parabolic and hyperbolic orbits, while $({\rm K}3)$ does not. This is why Gauss emphasized $({\rm K}3')$ by placing it, together with equation $(P)$, in the first pages of his book \cite{Gau}. Perhaps the more general applicability of $({\rm K}3')$ and $(P')$ may convince our readers that Gauss's message deserves their consideration.

\bigskip

\centerline{\bf 3. Kepler-Jacobi looks like Kepler}

\bigskip

Let us consider the {\it attractive case} of the Kepler-Jacobi problem: the acceleration vector is centripetal, that is, directed toward the Sun. This is a hypothesis on the sign of the coefficient $\rho^{\{2\}}$ in $(N')$. We assume more precisely that $\rho^{\{2\}}\geq 0$, and that $\rho^{\{2\}}=0$ on a finite number of rays in the plane ${\rm O} xy$. The following properties, rather familiar in the usual Kepler problem, are also true in this case.

$({\rm A}1)$ Any nonradial orbit covers the whole boundary of a strictly convex domain which contains ${\rm O}$. It does not have self-intersections or positions with zero velocity.

$({\rm A}2)$ There are open sets of the phase space filled up with periodic orbits.

$({\rm A}3)$ Two distinct orbits have at most two intersections.

{\it These properties are not Euclidean, they are all about convexity.} By the expression $(H)$ of the Hessian matrix $\partial^2\rho$, a hypothesis on the sign of $\rho^{\{2\}}$ is a convexity condition on $\rho$. 

We call {\it homogeneously strictly convex} a positively homogeneous $\rho:\R^n\setminus\{{\rm O}\}\to \R$ of degree 1 which satisfies
(i) $\rho(q_1+q_2)<\rho(q_1)+\rho(q_2)$ for any pair $(q_1,q_2)\in(\R^n)^2$ of linearly independent vectors and
(ii) $0<\rho(-q)+\rho(q)$ for any $q\neq {\rm O}$.

This is a kind of strict convexity adapted to the homogeneity, often discussed when dealing with norms or Finsler metrics. Here $\rho$ is not necessarily a norm: we did not assume $\rho>0$ or $\rho(-q)=\rho(q)$. The triangle inequality (i) gives (iii) $\rho(\nu_1q_1+\nu_2q_2)<\nu_1\rho(q_1)+\nu_2\rho(q_2)$ for any $\nu_1>0$, $\nu_2>0$. If $\nu_1+\nu_2=1$, this is the usual strict convexity condition.

Remarkably, if $n\geq 2$, (i) implies (ii), without any continuity hypothesis. Indeed, (i) implies that $\rho$ is convex on any convex subset of $\R^n\setminus\{{\rm O}\}$. Being locally convex, $\rho$ is continuous
on $\R^n\setminus\{{\rm O}\}$. We may consider the right-hand side of (ii) as a limit  of right-hand sides of (i). This limit cannot be negative, since the left-hand side of (i) tends to zero. Consequently $\rho$ extends to a convex function on $\R^n$, whose value at ${\rm O}$ is zero. If $\rho(-q)+\rho(q)=0$ for some $q\neq {\rm O}$, then the restriction of $\rho$ to the vector line directed by $q$ is affine. But {\it if a convex function is affine when restricted to a whole line, it is affine with same slope when restricted to a parallel line}. This is the epigraph version of: {\it if a closed convex set contains a whole line and a point, it contains the parallel line passing through the point}. These affine restrictions contradict the strict inequality (iii). So (ii) is proved.

Arguments due to Minkowski give this strict Hahn-Banach theorem:
{\it Any positively homogeneous $\rho:\R^n\setminus\{{\rm O}\}\to \R$ of degree 1 which is homogeneously strictly convex is such that there is a linear form $\omega$ with $\rho-\omega>0$ on $\R^n\setminus\{{\rm O}\}$.}

A differentiable enough $\rho: {\rm O} xy\setminus\{{\rm O}\}\to \R$ in the attractive case satisfies (i), since according to $(H)$ the restriction of $\rho$ to any line which does not pass through ${\rm O}$ is strictly convex. Consequently $\rho$ satisfies (ii) and is homogeneously strictly convex. The equation of the orbits being $\rho=\alpha x+\beta y+p$, adding a linear form to $\rho$ does not change the set of orbits. So, by Hahn-Banach, we may assume without loss of generality that $\rho>0$ on ${\rm O} xy\setminus\{{\rm O}\}$. This gives $({\rm A}2)$ since $\rho-\alpha x-\beta y=p$ has a solution on any ray if $|\alpha|$ and $|\beta|$ are small enough and $p>0$. The domain in $({\rm A}1)$ is defined by $\rho-\alpha x-\beta y\leq p$. It is strictly convex as the sublevel set of a homogeneously strictly convex function. The curve $\rho-\alpha x-\beta y=p$ is the whole orbit, since in the plane the only convex sets with disconnected boundary have two parallel lines as boundary. To get $({\rm A}3)$, we subtract the equations $(P')$ of two orbits. We get the equation of a line which according to $({\rm A}1)$ cuts an orbit at most twice.

Property $({\rm A}2)$ and its relation with Hahn-Banach are the result of a joint work with A.J.\ Ure\~na. Using the Clairaut-Binet method for solving the Kepler problem and its perturbations, we translated it into ``if a positively forced linear oscillator possesses periodic solutions, it possesses positive solutions'' (see \cite{AlU}). This simple claim was apparently unknown, as well as the application of convexity and the Hahn-Banach theorem to such questions.

\bigskip
\centerline{\includegraphics[width=45mm]{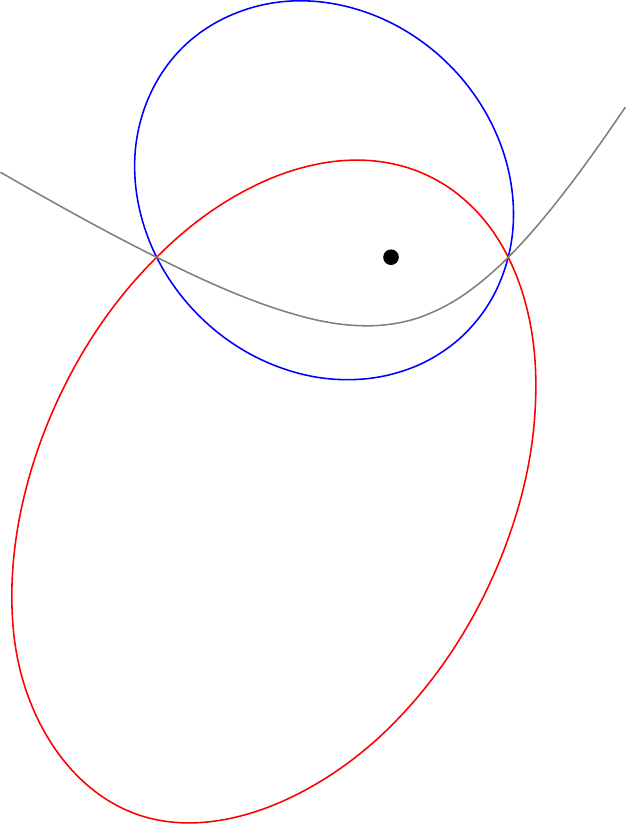}}

\nobreak
\centerline{Figure 2. Keplerian orbits with same angular momentum.}
\bigskip

Let us mention two properties of the Keplerian orbits passing through two points which are still there in the Kepler-Jacobi generalization.  {\it The absolute value of the angular momentum is the same on all the orbits passing through two given points of the ${\rm O} x$ axis, one with $x>0$, one with $x<0$} (see Figure 2). This is stated for example in \cite{God} as ``The magnitudes of the normal components at the terminals do not vary with the choice of trajectory.''

\bigskip
\centerline{\includegraphics[width=55mm]{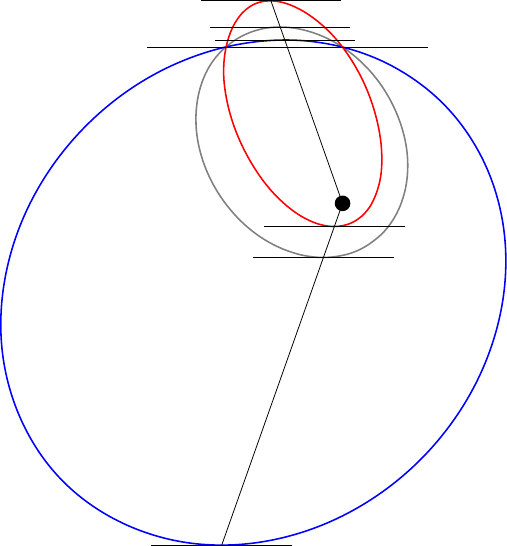}}

\nobreak
\centerline{Figure 3. Levine's property of Keplerian orbits with two common points.}
\bigskip

If the two common points are on another horizontal line (see Figure 3), the angular momentum depends on the orbit. Levine \cite{Lev}, while devising a ``method of orbital navigation using optical sightings,'' noticed that the direction from ${\rm O}$ to the point where $\dot y=0$ and $y>0$ does not depend on the orbit.

\bigskip

\centerline{\bf 4. The circular hodograph is modified in Jacobi's generalization}

\nobreak
\bigskip

The velocity vector $(\dot x,\dot y)$ moves on a curve called the hodograph. The hodograph of the elliptic orbit is a circle of radius $m|C|^{-1}$, as discovered by Hamilton in 1846 (see \cite{Ham} and Figure 4). The hodographs of the parabolic and hyperbolic orbits are circular arcs which end when the velocity vector is tangent to the circle.  To see this, just simplify the expression $(V)$ of the velocity 
$$\dot x=-\frac{m}{C}\Bigl(\frac{y}{r}-\beta\Bigr), \qquad \dot y=\frac{m}{C}\Bigl(\frac{x}{r}-\alpha\Bigr).\eqno(W)$$
On the right-hand sides, the first terms form a vector of norm $m|C|^{-1}$, the second terms a constant vector.

 \bigskip
\centerline{\includegraphics[width=80mm]{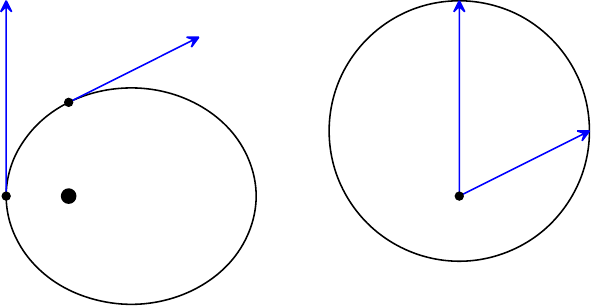}}

\nobreak
\centerline{Figure 4. A Keplerian orbit and its hodograph}
\bigskip

\centerline{\includegraphics[width=80mm]{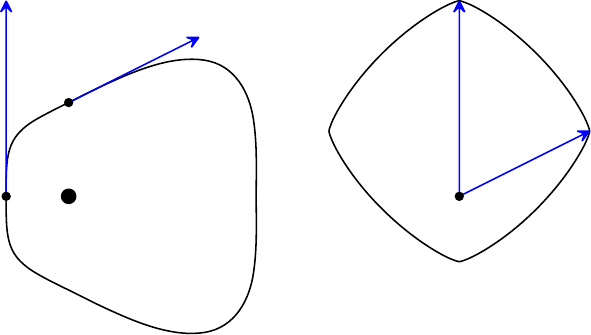}}

\nobreak
\centerline{Figure 5. An orbit for $\rho^4=x^4+y^4$ and its hodograph}
\bigskip

\centerline{\includegraphics[width=80mm]{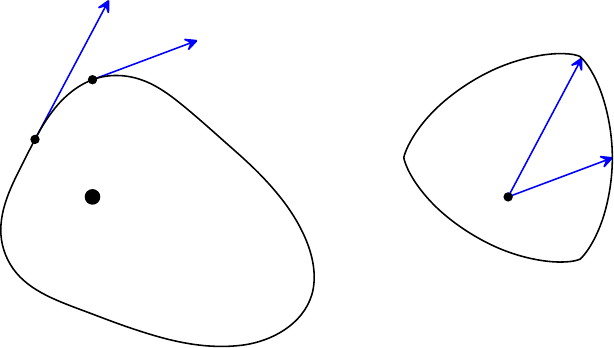}}

\nobreak
\centerline{Figure 6. An orbit for $\rho=2r+y^3/r^2$ and its hodograph}

\bigskip

The circle is a Euclidean figure. It should be replaced by something more general when passing to the Kepler-Jacobi problem, where a general $\rho$ replaces $r$. We get
$$\dot x=-\frac{m}{C}\Bigl(\frac{\partial \rho}{\partial y}-\beta\Bigr), \qquad \dot y=\frac{m}{C}\Bigl(\frac{\partial \rho}{\partial x}-\alpha\Bigr).\eqno(W')$$
The vector $(-{\partial \rho}/{\partial y},{\partial \rho}/{\partial x})$ is positively homogeneous of degree 0, that is, constant along the rays from ${\rm O}$. During the revolution of the ``planet,'' it moves along  a curve which is the same for all the orbits. The law of the circular hodograph is generalized: this unique curve replaces the circle. The hodograph of the orbit $\rho=\alpha x+\beta y+C^2/m$ is this unique curve translated by the vector $(\beta, -\alpha)$, rescaled by the factor $m/C$ and, in the unbounded case, truncated at points where the planet escapes to infinity.

\bigskip

\centerline{\bf 5. The energy first integral disappears in Jacobi's generalization}

\bigskip

In the Kepler-Jacobi problem with $\rho=\sqrt{Ex^2+2Fxy+Gy^2}$, where $(E,F,G)\in\R^3$ satisfy $E>0$, $EG>F^2$, the energy $$H=\frac{E\dot x^2+2F\dot x\dot y+G\dot y^2}{2}-\frac{m(EG-F^2)}{\sqrt{Ex^2+2Fxy+Gy^2}}$$ is constant along the orbits of $(N')$. The kinetic energy is expressed through the same quadratic form which defines $\rho$ and $\rho^{\{2\}}$. In contrast, in a Kepler-Jacobi problem where $\rho$ is not defined by a quadratic form, we cannot find anything which looks like a kinetic energy or an energy first integral.

In a domain where all the orbits are periodic, we could think of defining something as the energy through the period. The authentic Kepler's third law $n^2a^3=m$ does relate both quantities. The Keplerian ellipses with same semi major axis $a$ have same period $T=2\pi/n$. The classical relation between the energy and the semi major axis is
$$a=-\frac{m}{2H}.$$
But in the Kepler-Jacobi problem,  we had to change $({\rm K}3)$ into $({\rm K}3')$, that is, $n^2a^3=m$ into $C^2/p=m$. The new law is not about the period. It does not lead to any kind of energy.

We should recall another extension of the Kepler laws and the Newtonian attraction: the Euclidean space may be changed into a space of constant curvature. By coincidence, this extension also appeared in 1842, with a short note by Charles Graves \cite{Gra} which describes motion of a planet on a sphere.  In this generalization energy and semi major axis are well defined. Both are functions of the period (see \cite{Alt}).

\bigskip

{\bf Is there any physics in a Kepler-Jacobi problem?} There is little hope to give any physical meaning to a differential system $(N')$ for which the energy is not even defined. A gravitational or a Coulomb attraction cannot be anisotropic and homogeneous, but should rather be asymptotically isotropic at infinity. We may observe that a point source of light repels small particles with a central force in $1/r^2$. The anisotropy seems more conceivable for light than for gravitation, but we are still far from any realistic phenomenon.

\bigskip

{\bf Acknowledgements.} I wish to thank H.S.\ Dumas, G.F.\ Gronchi, M.\ Serrero, A.J.\ Ure\~na and the reviewers for their criticisms.

\bigskip

\centerline{\bf Appendix 1. Proof that Newton implies Kepler by the eccentricity vector.}

\bigskip

The classical expression of the eccentricity vector $(\alpha,\beta)$ in the planar Kepler problem is Equation $(W)$ in the form
$$\alpha=\frac{x}{r}-\frac{C\dot y}{m},\quad\beta=\frac{y}{r}+\frac{C\dot x}{m}\eqno(E)$$
This first integral of Newton's equation $(N)$ possesses a geometrical interpretation: it is a vector with Euclidean norm equal to the eccentricity, indicating the direction opposite the pericenter. The square of this norm is $$\alpha^2+\beta^2=\frac{x^2+y^2}{r^2}+\frac{2C(-x\dot y+y\dot x)}{mr}+\frac{C^2}{m^2}(\dot y^2+\dot x^2)=1+\frac{2C^2H}{m^2},\quad\hbox{with}\quad H=\frac{\dot x^2+\dot y^2}{2}-\frac{m}{r}.$$
Interestingly the energy $H$ appears as a factor of $\alpha^2+\beta^2-1$. In Jacobi's generalization, the eccentricity vector becomes the pair of first integrals $$\alpha=\frac{\partial \rho}{\partial x}-\frac{C\dot y}{m},\quad\beta=\frac{\partial \rho}{\partial y}+\frac{C\dot x}{m}.\eqno(E')$$
Here any Euclidean norm would be an arbitrary choice. Nothing like the energy first integral would appear as a factor.

To deduce that Kepler's laws are consequences of $(N)$, which means {\it solving} Newton's differential system $(N)$, it is enough to do the computation from \S 1 in the other direction. System $(N)$ clearly implies that $C=x\dot y-y\dot x$ has a zero time derivative. Integrating $(N)$ gives $(W)$ with the two constants of integration $\beta$ and $\alpha$. From $(W)$ we recompute $x\dot y-y\dot x$, which gives the focus-directrix equation in the form $C^2/m=r-\alpha x-\beta y$. So, Kepler's three laws are proved, the third one being $p=C^2/m$.

This method of integration is well known, close to J.\ Hermann's in 1710 \cite{Her}. Lagrange \cite{Lag} improved it by presenting the 3-dimensional analog in its final form in 1782 ---without vector notation. The key ingredient is the eccentricity vector. Hermann wrote the expression for one of its coordinates. Lagrange wrote all 3 coordinates. Surprisingly, this vector is sometimes called the LRL vector, for Laplace, Runge and Lenz, although none of these three authors ever claimed to have discovered it, while Hermann and Lagrange both have a legitimate novelty claim.

The method works with the Newton-Jacobi law in the 2-dimensional case, if $\rho$ is given. However, we should remark that what appears in the differential system $(N')$ is not the function $\rho$, but the function $\rho^{\{2\}}$.

Starting with a $\rho^{\{2\}}$, for example $(x^4+y^4)^{-3/4}$, can we find a $\rho$? This is generally not possible (see \cite{AlS}). A vector integral around a loop should vanish. However, the symmetry condition $\rho^{\{2\}}(x,y)=\rho^{\{2\}}(-x,-y)$, for all $(x,y)\neq (0,0)$, implies that the vector integral vanishes. It is satisfied if $\rho^{\{2\}}(x,y)=(x^4+y^4)^{-3/4}$. The expression for $\rho$ is then obtained by quadrature. In the example, quadrature gives the function $\rho$ in terms of Abelian integrals, associated to the genus 3 curve $x^4+y^4=w^4$.

\bigskip

\centerline{\bf Appendix 2. Is the Newton-Jacobi differential system Hamiltonian?}
\bigskip

As we cannot form an energy first integral in the general case $(N')$, we cannot put this system in the traditional Hamiltonian form:
The Newton-Jacobi dynamics is generally ``less'' Hamiltonian than the Newtonian dynamics. But it remains Hamiltonian in several ways. An anodyne remark is that the dynamics in the position space ${\rm O} xy$ admits the Hamiltonian $2\sqrt{m(\rho-\alpha x-\beta y)}$, if we fix $(\alpha,\beta)\in\R^2$. This follows immediately from $(W')$.

There is another way. If we accept changes of time and restrictions on the position, we can produce a Hamiltonian form by following a method described in \cite{Alt}.
We add an abstract coordinate $w$ such that the plane ${\rm O} xy$ is now the plane $w=1$ in the space $\Omega wxy$. We homogenize $(P')$ and for example the second formula $(E')$ into
$$\rho=\alpha x+\beta y+p w,\qquad E_\beta=\frac{\partial \rho}{\partial y}+\frac{(x\dot y-y\dot x)(w\dot x-x\dot w)}{m}.$$
If we fix $w=1$, $\dot w=0$, we re-obtain $(P')$ and $(E')$. But if instead we fix $x=1$, $\dot x=0$, the dot derivative now refers to a new time parameter, and we find the equation of orbits and the first integral in the form
$$\rho=\alpha+\beta y+p w,\qquad E_\beta=\frac{\partial \rho}{\partial y}-\frac{\dot y\dot w}{m}.$$
The Lagrangian, obtained by changing the minus sign of $E_\beta$ into a plus sign, gives the transformed equations of motion
$$\frac{\ddot w}{m}=\frac{\partial^2\rho}{\partial y^2}=\rho^{\{2\}},\qquad \frac{\ddot y}{m}=0.\eqno(L)$$
When we restrict the Kepler case to the plane $x=1$, we get $\rho=r=\sqrt{1+y^2}$ and $r^{\{2\}}=(1+y^2)^{-3/2}$.
This is the Kepler problem with the center of attraction sent to infinity by a projective transformation. It may be described as a Galilean law in the plane $(y,w)$ with a gravitational constant depending on the coordinate $y$. The Kepler-Jacobi case gives the same kind of law with a different dependence on $y$. These systems are Hamiltonian, since they come from a natural Lagrangian. We should indeed call them pseudo-natural since the kinetic energy $\dot y\dot w/m$ is indefinite.
The first formula $(E')$ together with the restriction $y=1$ gives a second Hamiltonian associated to a second change of time. The Newton-Jacobi system is consequently {\it quasi-bihamiltonian}, where {\it quasi} refers to the change of time. If instead of restricting to $y=1$ we continue the study of system $(L)$, we may extract a first integral from the first formula $(E')$. We simply restrict
$$E_\alpha=\frac{\partial \rho}{\partial x}-\frac{(x\dot y-y\dot x)(w\dot y-y\dot w)}{m}$$
to $x=1$, $\dot x=0$ and find
$$E_\alpha=\frac{\partial \rho}{\partial x}-\frac{\dot y(w\dot y-y\dot w)}{m}.$$
We may check directly that $\dot E_\alpha=0$. Let us repeat a message from \cite{Alt}: In a projective transformation, the transformed first integrals are easily computed. The transformed dynamical equations may be deduced from the transformed first integrals.

\bigskip

\centerline{\bf Appendix 3. Halphen's identity and Jacobi's attraction law.}

\bigskip

Let $n\in\N$. Let $g:\R\to\R$ be a differentiable enough function.  Denote by $g^{(n)}$ the $n$-th derivative of $g$. We have the following identity, where $^{(n)}$ in the left-hand side denotes the $n$-th derivative with respect to the variable $x$: $$\Bigl(x^{n-1}g\bigl(\frac{1}{x}\bigr)\Bigr)^{(n)}=\frac{(-1)^n}{x^{n+1}}g^{(n)}\bigl(\frac{1}{x}\bigr).$$

This surprising identity appears in some exercise lists for students. I found a report in \cite{Com}, p.\ 161, with a reference to Halphen and a proof using Lagrange's inversion formula. I propose a short proof. First note that this identity is immediately deduced from the following one. Recall our usual notation $\R^2={\rm O} xy$.

{\bf Proposition.} Let $n\in\N$ and $G:{\rm O} xy\setminus\{{\rm O}\}\to \R$ be a differentiable enough function, positively homogeneous of degree $n-1$. Then at any $(x,y)\in{\rm O} xy\setminus\{{\rm O}\}$
$$(-y)^{-n}\frac{\partial^n G}{\partial x^n}=x^{-n}\frac{\partial^n G}{\partial y^n}.$$

{\it Remark.} For $n=1$ this is  Euler's homogeneous function theorem for the degree 0. On the ${\rm O} x$ axis the left-hand side looks singular, but it is indeed continuous according to the right-hand side, if $x\neq 0$. The same occurs on the ${\rm O} y$ axis.

{\it Deduction of Halphen's identity.} Consider the function $G(x,y)=x^{n-1}g(y/x)$ which is positively homogeneous of degree $n-1$. Then
$$\frac{\partial^n G}{\partial x^n}=\frac{\partial^n}{\partial x^n}\Bigl(x^{n-1}g\bigl(\frac{y}{x}\bigr)\Bigr),\qquad \frac{\partial^n G}{\partial y^n}=\frac{1}{x}g^{(n)}\bigl(\frac{y}{x}\bigr).$$
We set $y=1$ and deduce Halphen's identity from the Proposition. Now, we should prove the Proposition and explain its relation with the Newton-Jacobi attraction.

To any sufficiently differentiable function $G:{\rm O} xy\setminus\{{\rm O}\}\to \R$ which is positively homogeneous of degree $n-1\in\N$ we associate its {\it scalar derivative} $G^{\{n\}}:{\rm O} xy\setminus\{{\rm O}\}\to \R$, positively homogeneous of degree $-n-1$, such that for any $k\leq n$, $$\frac{\partial^{n}G}{\partial x^k\partial y^{n-k}}=G^{\{n\}}|_{(x,y)}x^{n-k}(-y)^k.$$
This formula gives the Proposition. We deduce it exactly as we deduced $(H)$, which is the case $n=2$. The $n$-tensor $\partial^{n}G$ is symmetric and has the vector $(x,y)$ in its kernel since $\partial^{n-1}G$ is positively homogeneous of degree 0. Consequently, it is proportional to the tensor $$\pmatrix{-y\cr x}\otimes\cdots \otimes\pmatrix{-y\cr x}.$$ The scalar derivative $G^{\{n\}}$ is the proportionality factor. The function $\rho^{\{2\}}$ which is so important in the Kepler-Jacobi study now finds its general framework. It is the scalar derivative of the function $\rho$ which defines the $\rho$-focus-directrix curves.

\end{document}